\documentclass[prl,twocolumn,superscriptaddress]{revtex4}
\usepackage{dsfont}
\usepackage{graphicx,amsmath}
\def\be{\begin{eqnarray}}
\def\ee{\end{eqnarray}}
\def\bc{\begin{center}}
\def\ec{\end{center}}

\def\txst{\textstyle}
\def\rmF{{\rm F}}
\def\rmd{{\rm d}}
\def\om{\omega}

\newcommand{\lsim}{\stackrel{\scriptstyle <}{\phantom{}_{\sim}}}

\begin{document}
\title{
Spin excitonic and diffusive modes in superfluid Fermi liquids}
\author{E.E. Kolomeitsev}
\affiliation{University of Matej Bel, SK-97401 Banska Bystrica, Slovakia}
\author{D.N. Voskresensky}
\affiliation{National Research Nuclear University "MEPhI",Kashirskoe sh. 31, Moscow 115409, Russia}
\affiliation{GSI, Helmholtzzentrum f\"ur
Schwerionenforschung GmbH, Planckstrasse 1, 64291 Darmstadt, Germany}
%=================================================================
\begin{abstract}
A role of a  particle-particle p-wave spin interaction in Fermi liquids with s-wave pairing is
studied. Depending on the sign of the interaction  there arises  either the new exciton collective
mode below the pair-breaking threshold or the diffusive excitation mode above the threshold. The
Landau parameters which control the interaction strength are evaluated for various systems: the
dilute fermion gases, degenerate electron liquid, metals, atomic nuclei and neutron
matter. The interaction removes also the square-root singularity in the phase space of pair
breaking processes. It is shown how these effects  influence the neutrino emissivity in the
neutron Cooper-pair recombinations in neutron stars.
\end{abstract}
%\date{\today}
 \maketitle

%=======================================================================
%================   INTRODUCTION  ========================
Processes with recombinations of  Cooper pairs provide important information about interparticle
interactions and the pairing mechanisms in different fermionic systems: ordinary superconductors
\cite{Martin-69}, liquid $^3$He and $^3$He--$^4$He mixtures~\cite{VW90}, cold atomic gases
\cite{coldgas}, atomic nuclei \cite{FL}, neutron stars~\cite{MSTV90,NS}, and other systems. In
superconductors, they are studied by absorption of infrared radiation or by the Raman
scattering~\cite{Mon-Zawad-90}. In the cold fermion atom gas one can use the Stokes scattering to
detect the onset of the pairing~\cite{BB04}. Inverse pair breaking and formation (PBF) reactions
constitute an important mechanism of the neutron star cooling~\cite{FRS76,NS}. In these processes
the energy is released in the form of neutrino--antineutrino pairs radiating off the star.
Superburst ignition depth is sensitive to the value of the PBF emissivity in the inner neutron
star crust \cite{Cumming}. In ~\cite{Page} the PBF processes are suggested to be responsible for
the recently observed rapid cooling of the young neutron star in Cassiopeia A.
%======================

It was shown~\cite{collmode} that a residual interaction of single particle excitations, which
does not contribute to pairing, can bind them in a state orthogonal to the Cooper pair, generating
collective excitation modes in superconductors.
For superfluid $^3$He the similar mechanism was studied in \cite{Baldo}.
Interactions in the same spin channel, in which the pairing occurs, were studied so far.
The influence of the interaction in one spin channel on the pairing in
another channel has not yet been considered.
%=======================

In this Letter we study the effects of the p-wave interaction in the spin-one channel on
excitations in a Fermi system with the spin-zero  pairing. We calculate response induced
by the external spin- and helicity-density sources and show that depending on the sign of an
effective interaction there appears either a new exciton mode or a diffusive excitation mode. Then
we evaluate the strength of this effective interaction for different Fermi systems and, as an
example, calculate the neutrino emissivity in the PBF processes  for the neutron star with the
neutron pairing in $^1$S$_0$ state taking into account the effects of the  new collective modes and
correlations.

We use the Fermi-liquid theory approach extended to systems with pairing by Larkin and Migdal and
by Leggett~\cite{LM63}. For the processes induced by weak nucleon interactions this approach was
adopted in Ref.~\cite{KV08}. Interactions in particle-particle ($\xi$) and particle-hole ($\om$)
channels are essentially different. The interaction amplitude of two fermions with momenta
$\vec{p}=p_\rmF\, \vec{n}$ and $\vec{p}\,'=p_\rmF\, \vec{n}\,'$ before and after the interaction
in the $\xi$-channel is parameterized as
 $
\widehat{\Gamma}^\xi= \Gamma_0^\xi(\vec{n},\vec{n\,}')(i\sigma'_2)
(i\sigma_2) + \Gamma^\xi_1(\vec{n},\vec{n\,}')
(i\sigma'_2\,\vec{\sigma}\,') (\vec{\sigma}\,i\sigma_2)
 $
and in the $\om$-channel  as
 $
\widehat{\Gamma}^\om= \Gamma_0^\om(\vec{n},\vec{n\,}') {1}'\, {1}
+ \Gamma^\om_1(\vec{n},\vec{n\,}')\, \vec{\sigma}\,'\,
\vec{\sigma}\,.
 $
Here $p_\rmF$ stands for the Fermi momentum, $\vec{n}$ and  $\vec{n}{\,'}$ are  the unit vectors.
The unit matrices $1$ and $1'$ and the Pauli matrices $\vec{\sigma}$ and $\vec{\sigma}\,'$ act in the
nucleon spin space. Superscript "$\om$" indicates that the amplitude in this channel is taken for
$|\vec{q\,}\vec{v\,}_{{\rm F}}|\ll \om\ll \epsilon_{{\rm F}}$,  $v_\rmF$ is the Fermi velocity,
$\epsilon_{\rm F}$ is the Fermi energy and $q=(\om,\vec{q})$ is the transferred  4-momentum. The
coefficients of harmonic expansion of the scalar $\Gamma_0^{\xi,\om}$ and spin
$\Gamma_1^{\xi,\om}$ amplitudes, the Landau parameters, should be either evaluated microscopically
or extracted from analysis of the experimental data~\cite{FL}.

%============================================
The singlet pairing in the Fermi liquid  occurs owing to the attractive interaction, $a^2
\rho\,\Gamma_0^\xi={f_0^\xi}<0$. At zero temperature the paring gap $\Delta$ follows from equation
$-1/f_0^{\xi}=A_0/(a^2\,\rho)=\ln(2\, \epsilon_\rmF/\Delta)$, where $a$ is the pole residue and
$\rho$ is the density of states at the Fermi surface. This expression is naturally generalized for
finite temperature $T$, cf. Eq.~(5) in the second paper of Ref.~\cite{KV08}. Since  $g_0^\xi\equiv
0$  for the scattering of identical fermions, the spin interaction in the $\xi$-channel simplifies
as $a^2\rho\Gamma_1^\xi(\vec{n},\vec{n}')= g_1^\xi (\vec{n}\,\vec{n}')$. It is usually assumed
that the higher Legendre harmonics are much smaller~\cite{LM63}. Since we focus on the spin
channel, the interaction $\Gamma_{0}^\om$ decouples and can be dropped. In the $\om$-channel,
$a^2\rho\Gamma_{1}^\om= g_0^\om+g_1^\om (\vec{n}'\vec{n})$.
Contributions from the zeroth harmonics,
$g_0^\om$, are  accompanied by the factor $v_\rmF^2$, see Ref.~\cite{KV08}, and for
non-relativistic Fermi liquids under consideration (for $v_\rmF^2 \ll 1$) can be put zero. Thus we
remain with only tree relevant Landau parameters $f_0^\xi <0$, $g_1^\xi$ and $g_1^\om$. Let us
first put $g_1^\om $ zero and demonstrate the influence of the interaction in the $\xi$-spin-one
channel, $g_1^\xi$, on the pairing effects in the $\xi$-spin-zero channel. Then we recover
dependence on $g_1^\om $.

%=======================================================================
%==================== VERTICES ========================
Consider now an external perturbation,
which couples to the spin density operator $\vec{s}(x)=\psi^\dag(x)\vec{\sigma}\psi(x)$ and  the
helicity density operator $h(x)=\psi^\dag(x)(\vec{\sigma}
\hat{\vec{p}}+\hat{\vec{p}}\vec{\sigma})\psi(x)/(2\,m)$, where $\psi$ is the spinor of a non-relativistic
fermion,
$\hat{\vec{p}}$ is the momentum operator, and $m$ is the mass of the free fermion.
From these quantities one can build the axial ($A$) fermion current
$\mathcal{J}^\mu=(h,\vec{s})$. The Fourier
transform of its bare components after the Fermi liquid renormalization becomes
$J^{\om,\mu}(\vec{n},q)=(\vec{\sigma}\vec{\tau}_{A,1}^{\,\om},\vec{\sigma}\tau_{A,0}^\om)$.
Here
$\tau_{A,0}^\om=e_A/a$ and $\vec{\tau}_{A,1}^{\,\om}
=e_A v_\rmF\vec{n}/a$ are $\om$-vertices,
$e_A$ is the effective charge of the quasiparticle.
For $\Gamma_1^\om=0$, that we now exploit, the in-medium vertices
are $\tau_{A,0}(\vec{n},q) = \tau^\om_{A,0}$ and
$\vec{\tau}_{A,1}(\vec{n},q) =\vec{\tau}^{\,\om}_{A,1}(\vec{n},q)$.
For  $\Gamma^\om \neq 0$ these vertices are modified~\cite{KV08}.
Additionally, in a system with pairing there arise new vertices responsible for the
PBF processes:
$\vec{\sigma}\widetilde{\tau}_{A,0}(\vec{n},q)$ and $\vec{\sigma}
\vec{\widetilde{\tau}}_{A,1}(\vec{n},q)$.  They follow from
the solution of the Larkin-Migdal equations \cite{LM63,KV08},
\begin{subequations}
\label{LME}
\be
\widetilde{\tau}_{A,0}(\vec{n},q) &=& -\frac{g_1^\xi}{a^2\, \rho}
\Big(\big\langle (\vec{n}\,\vec{n}')\, ( N(\vec{n}',q)+A_0)\,
\widetilde{\tau}_{A,0}(\vec{n}',q) \big\rangle_{\vec{n}'}
\nonumber\\
&+& \big\langle (\vec{n}\,\vec{n}')\, O(\vec{n}',q;-1)\,{\tau}_{A,0}^\om\big\rangle_{\vec{n}'} \Big)\,,
\label{LME:tt0}\\
\vec{\widetilde{\tau}}_{A,1}(\vec{n},q) &=&-
\frac{g_1^\xi}{a^2\, \rho}\Big( \big\langle (\vec{n}\,\vec{n}')\,
( N(\vec{n}',q)+A_0)\, \vec{\widetilde{\tau}}_{A,1}(\vec{n}',q)
\big\rangle_{\vec{n}'}
\nonumber\\
&+&
\big\langle (\vec{n}\,\vec{n}')\,O(\vec{n}',q;+1)\,\vec{\tau}_{A,1}^{\,\om}(\vec{n}')
\big\rangle_{\vec{n}'} \Big)\,.
\label{LME:tt1}
\ee
\end{subequations}
The brackets indicate the angular averaging
$\langle \dots\rangle_{\vec{n}}=\int\frac{\rmd \Omega_{\vec n}}{4\pi}\, (\dots)\,. $
The loop functions
$O(\vec{n},q;\pm 1)  = {\txst\frac12} a^2 \rho\,(z_+\pm z_-) g_T(\vec{n},\om,\vec{q}\,)$,
and
$N(\vec{n\,},q)  = a^2 \rho\,z_+\, z_-\,g_T(\vec{n},\om,\vec{q}\,)$
with $z_\pm=(\om\pm \vec{v}\,\vec{q}\,)/(2\, \Delta)$, and
the master function
\be
%&&
g_T(\vec{n},\om,\vec{q}\,)=
\Delta^2\intop_{-\infty}^{+\infty}\frac{\rmd\xi_p}{\epsilon_+\, \epsilon_-}
\Big[
\frac{E_-\,F_-}{\om^2-E_-^2} - \frac{E_+\,(1-F_+)}{\om^2 - E_+^2}
\Big]\,,
%\label{gT}
\nonumber\ee
where $E_\pm =\epsilon_+ \pm \epsilon_-$,  $F_\pm =f(\epsilon_-)-f(\epsilon_+)$,   $f(x)=1/(\exp(x/T)+1)$
and $\epsilon_\pm=[(\xi_p\pm \vec{v}\vec{q}\,)^2+\Delta^2]^{1/2}$.
The solution of Eq.~(\ref{LME}) is
\be
&&\widetilde{\tau}_{A,0} =
-{\txst\frac{(\vec{v}\,\vec{q}\,)}{2\, \Delta}}\,{\tau}_{A,0}^\om\,\gamma^\xi_\parallel
\langle g_T(\vec{n}')\,(\vec{n}_q\,\vec{n}')^2\rangle_{\vec{n}'}\,,\quad
\label{taus}\\
&&\vec{\widetilde{\tau}}_{A,1}= {\txst \frac{\om}{q}} \vec{n}_q\widetilde{\tau}_{A,0}-
{\txst \frac{\om \, \tau_{A,1}^{\,\om}}{2\, \Delta}} \gamma_{\perp}^\xi
\langle g_T(\vec{n}'){\txst\frac12}[1-(\vec{n}_q\,\vec{n}')^2]\rangle_{\vec{n}'}
\vec{P}_\perp\,,
\nonumber
 \ee
where $\vec{P}_\perp=\vec{n}-\vec{n}_q\, (\vec{n}\, \vec{n}_q)$, $\vec{n}_q =\vec{q}/|\vec{q}|$
and the correlation factors
\be
[\gamma^\xi_\perp]^{-1}&=&
{\txst \frac13}\, C_0
+
\big\langle {\txst\frac{\om^2-(\vec{v}\,\vec{q}\,)^2}{4\, \Delta^2}} g_T(\vec{n})
{\txst \frac12}[1-(\vec{n}\vec{n}_q)^2]\big\rangle_{\vec{n}}\,,
\nonumber\\
{}[\gamma^\xi_\parallel]^{-1} &=&
 {\txst \frac13}\, C_0+
\big\langle {\txst\frac{\om^2-(\vec{v}\,\vec{q}\,)^2}{4\, \Delta^2}} g_T(\vec{n})
(\vec{n}\vec{n}_q)^2\big\rangle_{\vec{n}}\,
 \ee
are controlled by one effective interaction parameter
 \be
 C_0={3}/{g_1^\xi}-{1}/{f_0^\xi}.
 \ee
%=======================================================================
The singlet pairing occurs for $f_0^\xi <0$ and $3f_0^\xi <g_1^\xi$. Then, if $g_1^\xi <0$, we
have $C_0<0$, otherwise the p-wave pairing is preferable. For $g_1^\xi>0$ we have  $C_0>0$.

The response of
the Fermi system to the excitation ($A$) is determined by the symmetrical current-current
correlator $\Pi^{\mu\nu}(q)= {\txst\frac12}\langle {\rm Tr}\{ J^{\om,\mu}(\vec{n},q)\, J^{\nu}(\vec{n},q)\}
\rangle_{\vec{n}}$ with the in-medium current $J^\mu(\vec{n},q)=
\big(\vec{\sigma}\vec{\chi}_{A,1}(\vec{n},q),\vec{\sigma}\chi_{A,0}(\vec{n},q)\big)$ expressed via the
reduced current correlators derived in~\cite{KV08}:
\be
&&\chi_{A,0}(\vec{n},q)\!\!=\!\! L(\vec{n},q;-1)\,
\tau_{A,0}(\vec{n},q)
+M(\vec{n},q)\, \widetilde{\tau}_{A,0}(\vec{n},q), \nonumber\\
&&\vec{\chi}_{A,1}(\vec{n},q)\!\!=\!\! L(\vec{n},q;+1)\,
\vec{\tau}_{A,1}(\vec{n},q)
+M(\vec{n},q)\,
\vec{\widetilde{\tau}}_{A,1}(\vec{n},q),
\nonumber
\label{chis}
\ee
where
$
M(\vec{n},q) = -a^2\,\rho\,z_+\,g_T(\vec{n},\om,\vec{q}\,),
$
and
$\frac{L(\vec{n\,},q;\pm 1)}{a^2\, \rho} =
\big(\frac{z_+}{z_-}-1\big) g_T(\vec{n},(\vec{v}\vec{q}),\vec{q}\,)
-\big(\frac{z_+}{z_-}-{\txst\frac{1\mp 1}{2}}\big)\,g_T(\vec{n},\om,\vec{q}\,)
$\,.
The temporal and spatial  components of the tensor are
$\Pi^{00}=\langle \vec{\tau}_{A,1}^{\,\om}\, \vec{\chi}_{A,1}(\vec{n},q) \rangle_{\vec{n}}$,
and $\Pi^{ij}=\delta^{ij}\langle \tau_{A,0}^{\,\om}\, \chi_{A,0}(\vec{n},q)
\rangle_{\vec{n}}$ with
\be
&&{\txst \frac{1}{3}\sum_i}\Pi^{ii}= e_A^2 \rho\,
\Big\langle\frac{(\vec{v}\, \vec{q})}{\om- \vec{v}\,\vec{q}}
[g_T(\vec{n},(\vec{v}\vec{q}),\vec{q}\,)
-g_T(\vec{n},\om,\vec{q})]
\Big\rangle_{\vec{n}}
\nonumber\\
&&\qquad+
e_A^2 \rho\, \frac{v_\rmF^2\,\vec{q}^{\,2}}{4\, \Delta^2}\gamma^\xi_\parallel\,
\langle (\vec{n}_q\,\vec{n})^2\, g_T(\vec{n},\om,\vec{q})\rangle_{\vec{n}}^2\,,
\nonumber\\
%%%%%%%%%%%%%%%%%%%%%%%%%%%%%%%%%%%%%%%%%%%%%%%%%%%%%%%%%%%%%%%%%%%%%
&&\Pi^{00} = v_\rmF^2 {\txst\frac{1}{3}\sum_i}\Pi^{ii}  + {e_A^2\, \rho} v_\rmF^2
\langle g_T(\vec{n},\om,\vec{q})\rangle_{\vec{n}}
\\
&&\qquad+ {e_A^2 \rho} \,  v_\rmF^2\,\frac{\om^2}{2\, \Delta^2}
\, \gamma_{\perp}^\xi\,
\langle g_T(\vec{n},\om,\vec{q}){\txst\frac12}[1-(\vec{n}_q\,\vec{n})^2]\rangle_{\vec{n}}^2
\nonumber\\
&&\qquad +
{e_A^2 \rho} \,
v_\rmF^2\,\frac{\om^2-v^2_\rmF\, q^2}{4\, \Delta^2}\, \gamma_{\parallel}^\xi \,
\langle g_T(\vec{n},\om,\vec{q})\, (\vec{n}_q\,\vec{n})^2\rangle_{\vec{n}}^2.\nonumber
\label{Pis}
\ee
The mixed components are
$\Pi^{i0}=\Pi^{0i}=\vec{n}_q^{\,i}{\txst \frac{\om}{3|\vec{q}|}}\sum_j \Pi^{jj}$\,.

%==================== EXCITON MODES ========================
From Eq.~(\ref{taus}) we see that the external perturbation can induce a singular response in the
PBF amplitudes at the values $\om$ and $\vec{q}$ corresponding to the poles of the
functions $\gamma^\xi_\perp$ and $\gamma^\xi_\parallel$. These poles
determine the
new transverse and longitudinal collective modes (spin excitons). For
$\vec{q}=0$,  the longitudinal and transverse modes coincide and their frequency $\om$
follows from the condition
\be
C_0 + y^2 \Re \tilde{g}_T (y) =0\,, \quad y={\om}/({2\Delta}),\label{exciton-eq}
\ee
where  $\tilde{g}_T(y)\equiv g_T(0,2\Delta y-i\,0,0)$\,.

Although the full inclusion of the  $g_1^\om$-dependence
is rather tedious, the modification of Eq.~(\ref{exciton-eq}) is simply given  by
the replacement $\Re\tilde{g}_T(y)\to \Re\tilde{g}_T(y)/(1+\frac13\, g_1^\om\, \Re\tilde{g}_T(y))$\,.
For $|C_0|\gg 1$ it induces the shift
\be
C_0\to C=C_0 /(1+ C_0 g_1^\om/3 ).
 \ee
This relation interpolates between the limits $|C_0|\ll 3/|g_1^\om|$ when $C\approx C_0$,
 and $|C_0|\gg 3/|g_1^\om|$  when $C\simeq (3/g_1^\om)(1-3/(g_1^\om C_0))$.
So, parameter $C$  controls effects of residual interactions on the PBF processes.

%============================== FIG 1 =======================================
\begin{figure}
\centerline{\includegraphics[width=4.cm]{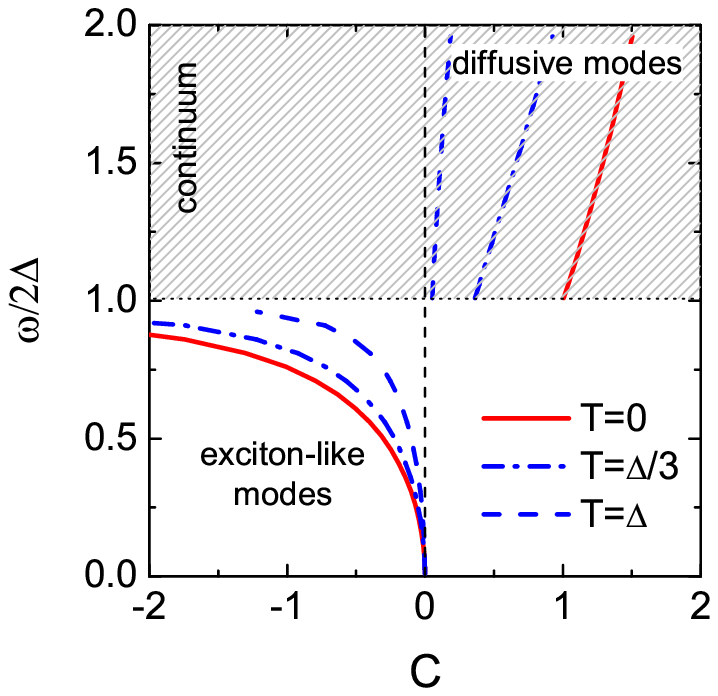}
\includegraphics[width=4.cm]{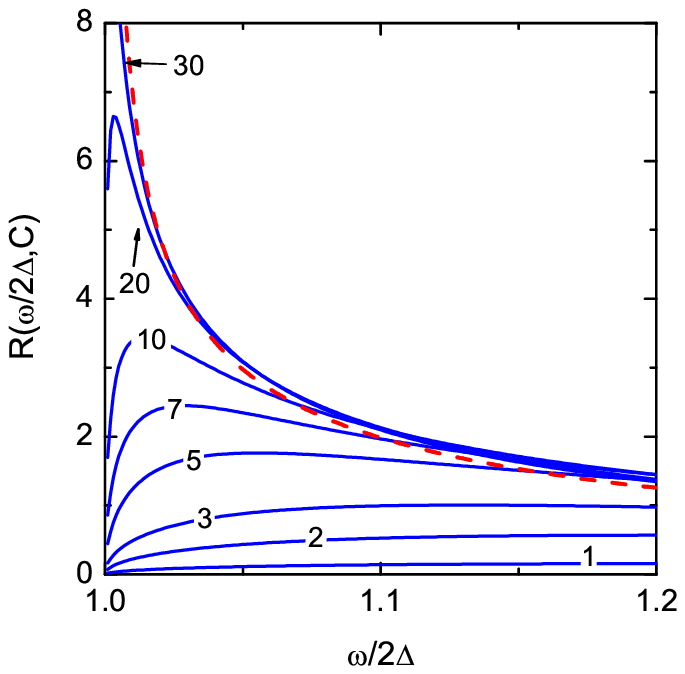}}
\caption{
Left: energies of the collective modes (for $\vec{q}=0$)
as functions of the parameter $C$ for various values of $\Delta/T$.
Right:  the response function $R(y)$ for $T=0$ and $y>1$ given by Eq. (\ref{Ffun}) for various
values of the parameter $C$ (solid lines). Dashed line shows the function $R(y, C\to \infty)$\,.
}
\label{fig:exciton}
 \end{figure}
%===================================================================================

%=========================CORRELATORS=======================================
In the long wave-length limit (for $\om>|\vec{q}|$) from Eq. (\ref{Pis})
we get
$ \Im\Pi^{ij}(q) = \frac{\delta^{ij}}{3} \frac{\vec{q}^{\,2}}{\om^2}
\Im\Pi^{00}(\om)$. The response function, having for $y\sim 1$ the form
\be
R(y,C)\equiv \frac{\Im\Pi^{00}}{e_A^2 \rho v_\rmF^2}=\frac{ C^2\Im \tilde{g}_T(y)
}
{[C+y^2 \Re\tilde{g}_T(y)]^2 +[y^2\Im \tilde{g}_T(y)]^2
}\phantom{xx}\label{Ffun}\\
+\pi\frac{C^2}{y^{2}}
\delta(C+y^2 \Re\tilde{g}_T(y)),\,\,
\Im \tilde{g}_T(y)=\frac{-\pi \tanh\big({\txst\frac{y\Delta}{2T}}\big) \theta(y)}{2y\sqrt{y^2-1}}
,
\nonumber
 \ee
determines the probability of  PBF processes. The cross section of the excitation scattering in
matter is determined by this response function $R$.

Solutions of Eq. (\ref{exciton-eq}) are shown in Fig.~\ref{fig:exciton} (left) as a function of
$C$. For $C<0$,  solutions with $y<1$ correspond  to the undamped spin exciton branch at  $\om
<2\Delta$, since $\Im \tilde g_T (y<1 )=0$. For $C>0$, solutions exist only if $C>-\Re
\tilde{g}_T(1+0)$ and $\om >2\Delta$. Since here $\Im \tilde g_T (y )\neq 0$, they constitute
the diffusive spin mode. The frequency of the exciton mode increases with the increase of $T$ and
decreases for the diffusive mode. The response function $R(y)$ at $T=0$ and $y>1$ is plotted in
Fig.~\ref{fig:exciton} (right panel) by solid curves  for  various  values of the parameter $C$.
For $y>1$ the function $R(y>1,C)$ only weakly depends on the sign of the value $C$ therefore we
present it only for $C>0$. The function $R(y,C\to \infty)=\Im\tilde{g}_T(y)$ is shown by the
dashed curve. It has a square-root divergence at $y\to 1+0$, which is smeared
out  for finite values of $C$.
Thus the finite value of $C$ leads to a reduction of the spin response of the Fermi liquid
close to the threshold for $\om>2\Delta$.
 A similar effect was discussed in Ref.~\cite{Mon-Zawad-90} for the
Raman scattering on metallic superconductors.

%==============================================================================
To estimate the value of our key parameter $C$ we need to know Landau parameters
$g_1^\xi$, $f_0^\xi$ and $g_1^\om$. For a dilute Fermi gas we can use
quasi-particle amplitudes derived in~\cite{nonidgas} up to the second order in the parameter
$\zeta=2 a_{\rm eff} p_\rmF/\pi$, where $a_{\rm eff}$ is the effective scattering length.
We derive
$f_0^\xi=\zeta+\zeta^2\,(2\ln 2+1)/3$, $g_1^\xi=3\zeta^2 (1-2\ln 2)/5$,
$g_1^\om=-2\zeta^2(\ln2+2)/5$ and obtain  $C\approx - 5.7/(a_{\rm eff} p_\rmF)^2$.
For the neutron gas the value of the $nn$-scattering length is very large, $a\simeq 20$ fm,
whereas the effective scattering length is much shorter \cite{SFB03}, being determined, e.g., by the
$V_{{\rm low}-k}$ potential, $a_{\rm eff}\simeq 2$~fm.

For more complex systems the parameters in the $\xi$-channel can be estimated with the help of the
$\om$-Landau parameters in the $s$--$p$ approximation of ~\cite{Patton75}
\be
f_0^\xi =\sum_{l=0}^\infty(-1)^l
\frac{A^s_l -3A^a_l}{4}\,,\,\,\,
g_1^\xi =\sum_{l=0}^\infty(-1)^l
\frac{A^s_l + A^a_l}{4}\,,
\label{f0xig1xi}
\ee
here $A^s_l={f^\om_l}/({1+\frac{f_l^\om}{2\, l+1}})$ and $A^a_l={g^\om_l}/({1+\frac{g_l^\om}{2\,
l+1}})$.

The Fermi-liquid approach was applied to the degenerate electron liquid in
Ref.~\cite{Kuechenhoff91}. Using Table~I and Table~II of ~\cite{Kuechenhoff91} we find $C=-2.54$, e.g.,
for a small value of the  parameter $a_{\rm B}p_\rmF=0.032$, where $a_{\rm B}$ is the Bohr radius.

For alkali metals at zero pressure the first three $\om$-harmonics  are calculated in
Ref.~\cite{Leiro95}. Applying (\ref{f0xig1xi}) we then find for  sodium
$f_0^\xi({\rm Na})=-0.11$, $g_1^\xi({\rm Na})=-0.38$ and $g_1^\om({\rm Na})=-0.075$. Here the p-wave
paring is realized, since $C_0 >0$, but the value $|C_0|$ is very small. Bearing in mind large uncertainties
in  estimates of the $\om$-Landau parameters one cannot exclude that $C<0$ at $|C|\ll 1$. In the latter case
we would deal with very pronounced effects of the spin exciton mode.
This case can also be realized, if one allows a variation of the pressure.
Thus presence or absence of the new exciton mode could tell about the kind of pairing in the given system.
For potassium $f_0^\xi({\rm K})=-0.56$,
$g_1^\xi({\rm K})=-0.89$ and, using $g_1^\om({\rm K})=-0.12$, we obtain $C=-1.48$.

\begin{figure}
\centerline{\includegraphics[width=7.5cm]{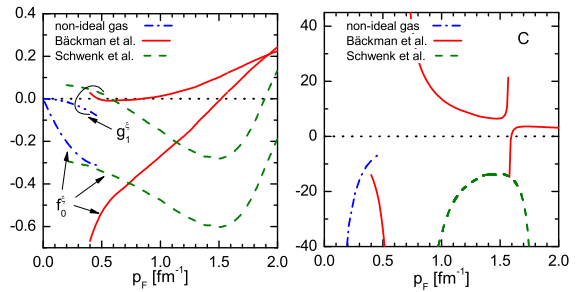}}
\caption{Parameters $f_0^\xi$ and $g_1^\xi$ (left panel) and   $C$ (right panel)
for the neutron matter
reconstructed with the $\om$-Landau parameters
calculated in Ref.~\cite{Backman73,SFB03} using  Eq.~(\ref{f0xig1xi})  as functions of the Fermi momentum.
}
\label{fig:LPnm}
\end{figure}

For the nucleon matter several harmonics of the $\om$-Landau parameters  were evaluated in many
works, e.g., see Refs.~\cite{Backman73,SFB03}. The parameter ${f_0^\xi}$ related to the $1S_0$
pairing was also calculated, see ~\cite{NS}. Contrary, the $g_1^\xi$ parameter is  poorly known.
Using results ~\cite{Backman73,SFB03} we  reconstruct $g_1^\xi$ and ${f_0^\xi}$ with the help of
Eqs.~(\ref{f0xig1xi}) and evaluate then parameters $C_0$ and $C$. For the neutron matter the
results are shown in Fig.~\ref{fig:LPnm} in dependence of the Fermi momentum. We see that
estimations of $C$ are very uncertain due to discrepancy in different estimates of the
$\om$-Landau parameters. Presented results show  that might be $|C|<10$--$20$ at some densities
in the  range of the 1S$_0$ paring and even $C$ might cross zero. Existence of regions where $C<0$
implies a possibility to observe effects of the exciton modes.

Using the values of the $\om$-Landau parameters and their density dependence extracted from the
atomic nuclear experiments \cite{FL,Borzov96}, we obtain $C\sim -10$ for $p_{\rm
F}\lsim 1$fm$^{-1}$. Thus the exciton mode could manifest itself in the nuclear surface phenomena.
%========================FIG 3 =======================================
%==============================================================================

%==================== THE NS COOLING ==========================================
Now we apply Eq.~(\ref{Pis}) to calculate the neutrino emissivity in the neutron star matter in
the region of $^1$S$_0$ pairing. It is mainly determined by the neutron PBF process
induced by the axial-vector current $\propto \mathcal{J}^\mu$~\cite{KV08}; the vector current
contribution is $O(v_{\rm F}^4)$ and can be neglected~\cite{Leinson,KV08}. For one type of
neutrino the emissivity then is given by \cite{KV08}
 \be
\varepsilon_{\nu\bar\nu} = G^2  g_A^2
\int^\infty_0\!\!\!\!\rmd \om  \om\!\int_0^\om\!\!\frac{\rmd |\vec{q\,}| \vec{q}^{\,2}}{48\, \pi^4}
\frac{(q_\mu\, q_\nu -g_{\mu\nu})\, \Im\Pi^{\mu\nu}(q)}{\exp(\om/T)-1},
\nonumber \ee
where $G$ and $g_A$ are the weak-interaction and axial-vector coupling
constants. The integration over $|\vec{q}\,|$
yields
\be
\varepsilon_{\nu\bar\nu}\simeq \frac{8}{35\, \pi^3} G^2 g_A^{2}e_A^2\rho\, v_\rmF^2 \Delta^7
\!\!\int^\infty_1\!\!\frac{\rmd y\, y^6 \, R(y,C)}{\exp(2y\Delta/T)+1},
\label{epsnn}
\ee
where according to Eq.~(\ref{Ffun}) there can be two contributions to $\varepsilon_{\nu\bar\nu}$:
one, for arbitrary $C$, from the pair-breaking continuum with the diffusive modes
at $\om>2\, \Delta$ and the other one, for negative $C$, from the spin-exciton mode
with the frequency  $\om(\vec{q})$ at $0<\om (\vec{q}=0)<2\,\Delta$.
The later contribution is associated with the processes of breaking and formation of spin excitons.
In the limit $|C|\to \infty$ the collective mode contribution vanishes as $\propto 1/|C|$
and we recover the  result~\cite{KV08}, $\varepsilon_{\nu\bar\nu}^{(0)}$, which follows
from (\ref{epsnn}) after the replacement $R(y,C)\to R(y,C\to \infty)=\Im\tilde{g}_T(y)$.
%============================== FIG 2 =======================================
\begin{figure}
\centerline{\includegraphics[width=4cm]{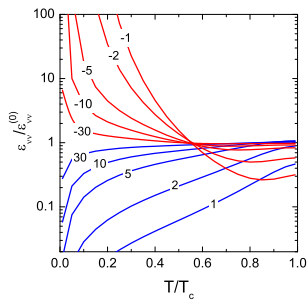}}
\caption{
The ratio $\varepsilon_{\nu\bar\nu}/\varepsilon_{\nu\bar\nu}^{(0)}$
as a function of $T/T_c$ for various values of $C$ (see curve labels). }
\label{fig:X}
\end{figure}
%==============================================================================

Effect of the finite value of $C$  on the neutrino emissivity in the neutron PBF process is
illustrated in Fig.~\ref{fig:X}, where we plot the ratio
$\varepsilon_{\nu\bar\nu}/\varepsilon_{\nu\bar\nu}^{(0)}$ taking into account the standard
temperature dependence of the $^1$S$_0$ pairing gap $\Delta(T)\simeq 3.1\, T_c\,(1-T/T_c)^{1/2}$
with $T_c$ as the critical temperature. For $|C|\sim 5$--$10$, cf. Fig. \ref{fig:LPnm} (right),
the effect becomes  pronounced for $T/T_c\lsim 0.5$, yielding a suppression for $C>0$ and an
enhancement for $C<0$. Thus in different density regions there may arise either an enhancement or
a suppression of the PBF emissivity. Effect becomes even more pronounced for smaller values  of
$|C|$.

In conclusion, we found that the spin p-wave interaction in the particle-particle channel can
produce new spin excitonic and diffusive modes in the Fermi system with the singlet paring. This
interaction leads also to smearing out of the threshold singularity in the Cooper-pair breaking
reactions. We calculated the relevant coupling parameters for several Fermi systems. Spin excitons
may exist in superconducting potassium, in rare fermion gases, and in the neutron matter. In
atomic nuclei the new spin exciton mode  may manifest in the  surface layer. Modification of the
neutrino emissivity  due to presence of spin excitonic and diffusive modes may have an impact on
the neutron star cooling.

We thank B.~Friman for discussions. The work was partially supported by ESF Research Networking
Programmes COMPSTAR and POLATOM, by the Alliance Program of the Helmholtz Association
(HAS216/EMMI), and by VEGA grant.

%================================================================================================


\begin{thebibliography}{99}
\bibitem{Martin-69}
P. C. Martin, in {\it Superconductivity}, ed. by R. D. Parks (Dekker, New York, 1969), Vol. I.

\bibitem{VW90}
D. Vollhardt, P. W\"olfle, {\it The Superfluid Phases of Helium 3} (Taylor, London 1990);
G.~Baym and C.J. Pethick, {\it Landau Fermi-Liquid Theory: Concepts and Applications} (Wiley, New York, 1991).

\bibitem{coldgas} S. Giorgini {\it et al.},\ Rev.\ Mod.\ Phys.\ {\bf 80}, 1215 (2008).

\bibitem{FL} A. B.~Migdal, {\it Theory of Finite Fermi Systems and Properties of
Atomic Nuclei}, Willey and Sons, N.Y. 1967.

\bibitem{MSTV90} A. B. Migdal {\it et al.}, Phys. Rept. {\bf 192}, 179 (1990).

\bibitem{NS} D. G.~Yakovlev {\it et al.}, Phys. Rept. {\bf 354}, 1 (2001).

\bibitem{Mon-Zawad-90} H. Monien and A. Zawadowski, Phys.\ Rev.\ B {\bf 41}, 8798 (1990).

\bibitem{BB04} G. M. Bruun and  G. Baym, Phys.\ Rev.\ Lett.\ {\bf 93}, 150403 (2004).

\bibitem{FRS76}
G.~Flowers {\it et al.}, Ap.\ J.\ {\bf 205}, 541 (1976);
D. N.~Voskresensky and A. V.~Senatorov, Sov.\ J.\ Nucl.\ Phys.\ {\bf 45}, 411 (1987).

\bibitem{Cumming} A. Cumming {\it et al.}, Ap. J. {\bf 646}, 429 (2006);
S. Gupta {\it et al.}, {\it ibid.} {\bf 662}, 1188 (2007).

\bibitem{Page} D.~Page {\it et al.}, Phys.\ Rev.\ Lett.\  {\bf 106}, 081101 (2011);
D. G.~Yakovlev {\it et al.},  Mon.\ Not.\ Roy.\ Astron.\ Soc.\  {\bf 411}, 1977 (2011).

\bibitem{collmode}
A. Bardasis and J. R. Schrieffer, Phys.\ Rev.\ {\bf 121}, 1050 (1961);
V. G. Vaks {\it et al.}, Sov.\ Phys.\ JETP {\bf 14}, 1177 (1962);
A. I.~Larkin, {\it ibid.}\ {\bf 19}, 1478 (1964);
P. Fulde and  S. Strassler, Phys.\ Rev.\ {\bf 140}, A519 (1965).

\bibitem{Baldo} M. Baldo et al., Phys. Lett. A {\bf 65}, 418 (1978).

\bibitem{LM63}
A. I.~Larkin and A. B.~Migdal, Sov.\ Phys.\ JETP {\bf 17}, 1146 (1963);
A. J.~Leggett, Phys.\ Rev.\  {\bf 140}, A1869 (1965); {\bf 147}, 119 (1966).

\bibitem{KV08} E. E.~Kolomeitsev and D. N.~Voskresensky, Phys.\ Rev.\ C {\bf 77}, 065808
(2008); {\it ibid} {\bf 81}, 065801 (2010).

\bibitem{nonidgas}
A. A.~Abrikosov and  I. M. Khalatnikov, Sov. Phys. JETP {\bf 6}, 888 (1958);
H. H.~Fu and Ch. J.~Pethick, Phys. Rev. B {\bf 14}, 3837 (1976);
M. Yu.~Kagan and A. V.~Chubukov, JETP Lett. {\bf 47}, 614 (1988).

\bibitem{SFB03} A.~Schwenk {\it et al.}, Nucl. Phys. A {\bf 713}, 191 (2003).

\bibitem{Patton75} B. R.~Patton and A.~Zaringhalam, Phys. Lett. A {\bf 55}, 95 (1975).

\bibitem{Kuechenhoff91} S.~K\"uchenhoff and S.~Schiller, Phys. Rev. B {\bf 43}, 10310 (1991).

\bibitem{Leiro95} J. A.~Leiro, Solid State Comm. {\bf 93}, 953 (1995).

\bibitem{Backman73} S.-O.~B\"ackman {\it et al.}, Phys. Lett. B {\bf 43}, 263 (1973).

\bibitem{Borzov96} I. N.~Borzov {\it et al.}, Z. Phys. A {\bf 355}, 117 (1996).

\bibitem{Leinson} L. B. Leinson and A. Perez, Phys. Lett. B {\bf 638}, 114 (2006).

\end{thebibliography}
\end{document}